----------
X-Sun-Data-Type: default
X-Sun-Data-Description: default
X-Sun-Data-Name: fpvcm.tex
X-Sun-Content-Lines: 989

\documentstyle[12pt]{article}
\def\eq{\begin{equation}}
\def\en{\end{equation}}
\newcommand \be  {\begin{equation}}
\newcommand \bea {\begin{eqnarray} \nonumber }
\newcommand \ee  {\end{equation}}
\newcommand \eea {\end{eqnarray}}

\def\id{\equiv}

\def\d{{\rm d}}
 \def\(({\left(}
 \def\)){\right)}
\def\bi{\bibitem}

\def \jij{J_{ij}}

\def \tp{\tilde{p}}
\def \ov{\over}
\def \a{\alpha}
\def \b{\beta}
\def\D{\Delta}
\def\bQ{{\bf Q}}
\def \d{{\rm d}}

\def \de{\delta}
\def \nn{\nonumber}
\def \beqna{\begin{eqnarray}}
\def \eeqna{\end{eqnarray}}
\def \beq{\begin{equation}}
\def \eeq{\end{equation}}
\def \be{\begin{equation}}
\def \ee{\end{equation}}

\def \ov{\over}

\def \a{\alpha}

\def \b{\beta}
\def \r{\right}

\def \l{\left}

\def \la{\langle}
\def \ra{\rangle}
\def \Tr{{\rm Tr}}
\def \eps{\epsilon}
\def \ab2{\alpha\beta^2}
\def \s{\sigma}
\def \la{\langle}
\def \ra{\rangle}

\newcommand \oo {\omega}

\begin{document}

\title{Interfaces and lower critical dimension in a spin glass model}
\author{ S. Franz (1), G. Parisi (2), M.A. Virasoro (2)\\
\\
 (1)  NORDITA, Blegdamsvej 17, DK-2100 Copenhagen \O, Denmark\\
(2) Universit\`a di Roma I "la Sapienza", P.le A. Moro 5, 00185 Rome Italy}

\maketitle

\begin{abstract}
In this paper we try to estimate the lower critical dimension
for replica symmetry breaking in  spin glasses through the calculation
of the additional free-energy required to create a domain wall between
two different phases.
This mechanism alone would say that replica symmetry would be restored
at the lower critical dimension $D_c =2.5$.

\end{abstract}
\vfill
%{\bf \hfill ROM}
cond-mat/9405007
\par\noindent
NORDITA preprint 94/21
\vfill
\newpage

The calculation of free energy increase  due to an interfaces is a
well known method to obtain information about
the lower critical dimension for spontaneous symmetry breaking.
We perform the first analytic
computation of this free energy increase in spin glasses and we use it to
suggest the value of the lower critical dimension.

Let us introduce  the basic definitions. In the simplest case we can
consider a system with two possible coexisting phases ($A$ and $B$).
We will study what happens in a finite system in dimensions $D$ of size
$M^d \  L$ with  $d=D-1$.  Let us  assume that in the $d$
transverse direction we have periodic or free
 boundary conditions, while in the
other direction (which we call $t$), we put the system in phase $A$ at
$t=0$ and in phase $B$ at  $t=L$. The free energy of the interface is the
increase in free energy due to this choice of boundary conditions with
respect to choosing the same phase at $t=0$ and $t=L$ (for simplicity
we suppose that the two choices $A$ or $B$ lead to the same value for
the free energy).

In many cases we have that the free energy increase $\delta F(L,T)$
behaves for large $M$ and $L$ as:
\be
\delta F(L,M)= M^d/L^\oo
\ee
where $\oo$ is independent from the dimension. There is then a critical
dimension at which the free energy of the interface is finite:
 \be
D_{c}=\oo+1.
\ee
 Heuristic
arguments \cite{parisibook1}, which sometimes can be made rigourous ,
tell us that at this dimension (the lowest critical dimension)
the two phases mix in such a way that symmetry is restored.

In most cases the value of $\oo$  from mean field theory is the exact
one and therefore we can calculate in this way the value of the
lower  critical dimension. The simplest examples are the
ferromagnetic  Ising model $\oo=0$ and the ferromagnetic Heisenberg
model $\oo=1$.

In this note we study the problem for spin glasses. For convenience we
consider two replicas of the same system described by a total hamiltonian:
\be
H=H[\s]+H[s]
\ee
where $H$ is the Hamiltonian of a short range spin glass.

 In this case the order parameter is the local overlap $q_i=\la\s_i s_i\ra$
between the two replicas. In mean field theory one finds that
this overlap is constant in space, and all values
in the  interval $[q_{min},q_{max}]$ are possible. A perturbative
expansion  around the solution with $q$ constant in space can be done in
sufficiently  high dimensions, while infrared divergences appear in
low dimension.

The aim of this note is to compute the free energy increase corresponding
to imposing an expectation value of $q$ equal to $p_1$ at
$t=0$ and $p_2$ at $t=L$. We consider here only the case where both
$p$'s and non zero and are in the interval $[q_{min},q_{max}]$.

We find that
\be
\delta F= V (|p_1-p_2|/L)^{5/2},
\ee
with $V\equiv M^d L$. As a consequence, the {\sl naive}
prediction of mean field theory for the lower critical dimension for
spontaneous replica symmetry breaking is $D_c=2.5$. If a similar value would
have been obtained for the interface free energy
 in zero magnetic field and
 with $p_1$ and $p_2$
of opposite sign, one would argue that the EA parameter should
vanish at $D=2.5$.

We stress that these predictions are {\sl naive}; corrections to the mean
field theory are neglected. While in ferromagnets there is regime (low
temperature) at which these corrections can be shown to be small,
in the spin glass case the corrections
to the mean field theory do not vanish even  at zero temperature.
We do not have here a regime in which we can show that the corrections to the
free energy interface do not change qualitatively its $L$ dependence.

A simple testable prediction of our computation
is that on a $L^D$ system with
boundary condition $q=q_{EA}$, the expectation value of $q$ in the
centre of the box should go for large $L$ as
\be
q(L)=q_{EA}- const/L^{(2D/5-1)}
\ee

Our paper is organized as follows. In section 1 we briefly outline
the results of the mean field theory for short range spin glasses
and we discuss its extension to the case of two constrained real replicas.
In section 2 we show that equation (4) can be derived using rather general
scaling arguments
under the technical assumption that the overlap between the
two constrained replicas varies linearly in space.
A confirmation for that behaviour is found
in section 3 where we find a variational approximation to the
free energy increment $\delta F$. In the final section we draw some
conclusions.

\section{The model}
The model we consider is
the standard D-dimensional Edwards-Anderson spin-glass \cite{EA} on a
square lattice in a finite volume $V$, which for simplicity will be taken as a
box of size $L>>1$. This is defined by the
Hamiltonian:
\be
H[s_i]=-\sum_{<ij>}J_{ij} s_is_j-h\sum_{i}s_i
\ee
where with standard notations we have denoted by  $<i,j>$  the nearest
neighbours
on the lattice. The spins are Ising variables  $s_i=\pm 1$, and
the couplings $J_{ij}$ are
independent  gaussian variables with zero average and
variance $\bar{\jij^2}=J^2=1$. $h$ is the magnetic field.

In our discussion we will make extensive use of the results of the mean field
theory (MFT)
of the model near the critical temperature. Let us summarize here, without
derivation, the
main results. For a more complete exposition of the theory
see e.g. \cite{mpv,parisibook2,binyou,fishertz}.

The study of the equilibrium properties of the model can be performed
 in the frame of replica method.
The relevant order parameter is a space dependent
overlap matrix $ Q_i^{ab}=\la  s_i^a  s_i^b \ra $ which, analogously
to the long range case, describes the statistics of
overlaps between pure states.
In MFT, where the system is treated in the saddle point
approximation, one finds
a (de Almeida-Thouless) line of
second order phase transition to a glassy phase, which terminates for $h=0$
in $T_c=1$.
In the vicinity of $T_c$, the free
energy as a functional of the order parameter admits a Landau expansion
in which the original lattice is coarse grained, and one considers the
 order parameter averaged
in small regions of the space $V_x$ centered in $x$, $Q_{ab}(x)={1\ov |V_x|}
\sum_{x\in V_x} Q_i^{ab}$.
The free energy functional  in terms of $Q_{ab}(x)$ is then written as:
\beqna
&- 2 n F=\int \d^D x[ {1\ov 2}\Tr Q(x)\Delta Q(x) +\tau \Tr Q^2(x) +(1/3) \Tr
Q^3(x)
\nn\\
&+(y/4) \sum_{ab}Q^4_{ab}(x)+h^2\sum_{ab}Q_{ab}(x)]
\label{landau}
\eeqna
where $\Delta$ is the Laplacian operator, $\tau=T_c-T=1-T$, $y=2/3$.
The integration extends to the square box of size $L$, and 'Tr' denotes the
trace in replica space.
As usual, among all the quartic terms in $Q$ which should be written  in the
expansion,
 we have only included the one
responsible for the de Almeida-Thouless instability \cite{par2}.
This  leads to
the phase transition into a replica symmetry breaking phase.
%It has been shown \cite{marcprl} that the other
%quartic terms present do not affect the physics in the vicinity of the
%transition (see however \cite{kondor}).

The main advantage of this reduced
model is that it allows for a complete analysis
of the r.s.b.: the saddle point equations can be solved exactly above and
below the transition temperature.
In the low temperature phase,
the solution to the saddle point equations is found
in the framework of Parisi Ansatz \cite{par1},
which in the present context consists in
parametrizing the space dependent matrices $Q_{ab}(x)$ as a space dipendent
functions
$q(x,u)$ with $0\le u\le 1$.
For free boundary conditions, the relevant saddle point is found to be
constant in
space, and the analysis become identical to that of the long range
SK model
\cite{par2}.
At the saddle point one finds:
\be
q(x,u)=q(u)=
\l\{
\begin{array}{ll}
q_{min}& u\le u_0\\
{u\ov 3y}& u_0\le u\le u_1\\
q_{max}& u\ge u_1
\end{array}
\r.
\label{qfree}
\ee
with $u_0=3y q_{min}$, $u_1=3 yq_{max}$.  $q_{min}$ and  $q_{max}$ are
 specified by the relations
\beqna
&2 y q_{min}^3=h^2\nn\\
&\tau=q_{max}(1-{3y\ov 2} q_{max})
\eeqna
As in long range models, the appearence of
 r.s.b.
imply the existence of many pure states with a
non-trivial distribution of the mutual overlaps \cite{par3}.
The theory predicts that  these overlaps are
constant in space.

Let us now enter into the core of our discussion. We want to
introduce  boundary conditions in the model to force spatial
 dishomogeneity
of the order parameter. In analogy with what is done
in ordered systems we would like to put the system in two different
equilibrium states at the two boundaries along a given
direction. We observe that this can not be done by imposing boundary values to
the function $q(x,u)$. Any variation with respect to the form (\ref{qfree})
would  take
us outside of the saddle point, where the free energy functional has
no physical meaning. We shall follow instead
 a procedure introduced in a previous paper
\cite{fpv1}(referred as I in the following),
to study long range spin glasses off-equilibrium.

Consider two identical (same $J_{ij}$) replicas of the system,
which undergo different thermal histories (for equal temperature and
magnetic field).
We constrain the overlaps between these two real replicas
(RR in the following)
on the two boundaries along one direction and leave free boundary conditions
on all others. In this case the Saddle Point overlap will be constant
 in all but the privileged direction, and we will have to solve
an effective one dimensional problem. Denoting
 $\partial_1$ and $\partial_2$
the boundaries on which we impose the non trivial constraint, we can write
the partition function for the doubled  system
as
\beqna
Z_c&=&
\sum_{ \{ s_i,\sigma_i \} }
\exp
\l\{
-\b H[s] -\b H[\sigma]
\r\}
\prod_{x\in\partial_1}\delta
\l(
{1\ov |V_x|}\sum_{x\in V_x} s_i \sigma_i -p_1
\r)
\nn\\
&\cdot &
\prod_{x\in\partial_2}\delta
\l(
{1\ov |V_x|}\sum_{x\in V_x} s_i \sigma_i -p_2.
\r)
\eeqna
The object of interest will be
 the free energy difference $\delta F$ between this situation and the
unconstrained case where the delta functions are not present in the partition
sum. We will get an estimate of $\de F$ in mean field theory. This
will enable us to  estimate the probability of
 fluctuations of the overlap of amplitude $|p_1-p_2|$ over a scale $L$ through
$Prob(|p_1-p_2|,L)\sim \exp(-\beta \delta F)$.
If this probability remains finite  for large $L$, this kind of fluctuations
destroy replica symmetry breaking.
In high enough dimension we will find a free energy difference divergent
 with $L$. The critical dimension $D_c$ will be then
identified as the dimension at which this free energy difference become finite.

In the replica treatement of the problem, one has to replicate both
the $s$ and the
$\sigma$ spins \cite{fpv1,fpv2}. Thus
three space dependent $n\times n$  matrices will appear:
$Q_{ab}^{11}(x)$
describing the overlaps among $s$ spins,
$Q_{ab}^{22}(x)$
describing the overlaps among $\sigma$ spins and
$Q_{ab}^{12}(x)$
describing the overlaps between $s$ and $\sigma$ spins.
For symmetry reasons $Q_{ab}^{12}(x)=Q_{ba}^{21}(x)$; the diagonal
elements
$Q_{aa}$ are as usual taken to be zero by convention.
We will choose in the following  saddle points for which
$Q_{ab}^{11}(x)=Q_{ab}^{22}(x)\id Q_{ab}(x)$ and
$Q_{ab}^{12}(x)=Q_{ab}^{21}(x)\id P_{ab}(x)$.
The constraint introduced in the partition function reflects itself
in the order
parameter through the fixing of the values of the diagonal
elements of $Q_{ab}^{12}$
on the border:
\be
Q_{aa}^{12}(x)|_{x\in \partial_1}=p_1; Q_{aa}^{12}(x)|_{x\in \partial_2}=p_2.
\label{constr}
\ee
All other elements of the matrices  are to be determined from
the variational equations for  the free energy.
In dealing with the matrices $Q^{rs}_{ab}$ $r,s=1,2$ $a,b=1,...,n$ it is useful
to introduce
new indices $\alpha\id (r,a)$, $\beta\id (s,b)$ and a $2n\times 2n$ matrix
${\bf Q_{\a\b}}=Q^{rs}_{ab}$ that contains all the three matrices.
In term of this new  matrix, the free energy functional is
formally  identical to that for a single real replica in term of the usual
 $Q_{ab}$. The  difference lies in the variational procedure, where
 one has to keep into account the constraint (\ref{constr}).

Near the critical temperature, the free energy admits a Landau like expansion
(see \ref{landau}),with the single replica matrix  $Q$ substituted by
${\bf Q}$. The saddle points equations in terms of the matrices $Q$ and $P$
are:
\beqna
  \Delta Q_{ab}&=&2\tau Q_{ab} +(Q^2)_{ab}+(P^2)_{ab}+yQ_{ab}^3+h^2
\nn
\\
  \Delta P_{ab}&=&2\tau P_{ab} +2 (QP)_{ab}+yP_{ab}^3+h^2
\eeqna
As usual, to solve these equations we need an ansatz that will eventually
allow us to do the
analytical continuation to $n\to 0$.
As in I we assume both matrices $Q$ and  $P$ to be
hierarchical matrices. Namely, we parametrize
$Q_{ab}(x)$ by a function of $u\in [0,1]$ $q(x,u)$  and the matrix
$P_{ab}(x)$ by a diagonal element $P_{aa}(x)=\tilde{p}(x)$ and
a function $p(x,u)$.
Physically $\tp(x)$ represents the space dependent overlap between two RR
constrained on the boundary. A tentative discussion of the physical meaning
of the functions
$q(x,u)$ and $p(x,u)$ can be found in I.

Clearly, for the chosen boundary conditions,  the various
parameters will be constant
 in all but one direction. Labelling the
coordinate along this direction by $t$ we find that
$q$ $p$ and $\tp$ will depend on space only through $t$.
In this way, denoting  the integrals
of the kind $\int_0^1\d u\;g(t,u)$ as  $\la g\ra$,
 the saddle point equations become:
\beqna
  {\partial^2\ov \partial t^2}q(t,u)&=&2(\tau-\la q\ra ) q(t,u)+2(\tp-\la
p\ra)p(t,u)
\nn\\
&+&\int_0^u \d v\; [q(t,u)-q(t,v)]^2+\int_0^u \d v\; [p(t,u)-p(t,v)]^2+
y q^3(t,u)+h^2
\nn\\
  {\partial^2\ov \partial t^2}p(t,u)&=&2 (\tau-\la q\ra ) p(t,u)+2(\tp-\la
p\ra)q(t,u)
\label{eqt}\\
&+&2\int_0^u \d v\; [q(t,u)-q(t,v)][p(t,u)-p(t,v)]+
y p^3(t,u)+h^2
\nn\\
  {\partial^2\ov \partial t^2}\tp(t)&=&
2 \tau \tp(t) -2 \la qp\ra +y \tp(t)^3 +h^2
\nn
\eeqna
If $p_1=p_2$ and $q_{min}\le p_1\le q_{max}$
one can find a solution without
spatial dependence at all, with $\tp=p_1$.
Then the problem reduces to the one discussed in I,
There we
showed, on very general grounds, that
for any  $p_1$
in this interval, there exist a solution
 to the saddle point equation which has the same free-energy density
of the unconstrained system $(\de F=0)$.
The results of I in the present context
are:
\be
q(x,u)=q(u)=
\l\{
\begin{array}{ll}
q_{min}&  u\le u_0/2\\
2x/(3y) &  u_0/2< u \le u_p/2\\
\tp &  u_p/2<u \le u_p\\
x/3y & u_p <u\le u_1\\
q_{max} &   u_1<u\le 1.
\end{array}
\r.
\nn
\ee
\be
p(x,u)=p(u)=
\l\{
\begin{array}{ll}
q_{min}&  u\le u_0/2\\
2x/(3y) &  u_0/2< u \le u_p/2\\
\tp &  u_p/2<u \le 1\\
\end{array}
\r.
\label{s0}
\ee
\be
\tilde{p}(x)=\tp=p_1
\nn
\ee
The parameters $q_{min}$, $q_{max}$, $u_0$ and $u_1$ are those of the
unconstrained solution (\ref{qfree}) and $u_p=3 y p_1$ is the point
where the function (\ref{qfree}) is equal to $p_1$.
This solution reflects the fact that the multiplicity of states
does not give an extensive contribution to the free energy. The
space of
equilibrium states of the  two copies constrained at an overlap $p_1$ is
simply a restriction of the cartesian product of the equilibrium states of
two free system.
The freedom in the choice of $p_1$, which is a zero mode of the free energy,
is the spin glass analog to the Goldstone zero mode found in ordered models
with a spontaneously broken continuous symmetry.
It is now clear that if we impose $p_1\ne p_2$, but both in
the interval $[q_{min},q_{max}]$, we are choosing
at the boundaries two
of the possible overlaps admitted by the {\it free} problem. The additional
free energy will have to go to zero in the limit when the boundaries
become very far from each other. Our aim is to know with what power it
goes to zero.

\section{A dimensional argument.}

To study the situation $p_1\ne p_2$ we could perturb around the solution with
$p_1=p_2$.
Above the critical dimension the laplacian term in (\ref{landau})
can be treated as a small
perturbation. The relevant
parameter of the expansion turns out to be $|p_2-p_1|/L$ which is always
arbitrary small.

It is reasonable to think that the solution to the saddle point equations
would be in this case {\it similar} in form to eq.(\ref{s0}), but with
$\tilde{p}=\tilde{p}(t)$ a function interpolating between $p_1$ at
$t=0$ and $p_2$ at $t=L$.
 On physical
grounds we expect that after averaging over the quenched disorder,
 $\tp(t)$ interpolates linearly between the boundary
values, namely
\be
\tp(t)=p_1 (1-t/L)+p_2 t/L.
\label{LI}
\ee
This assumption
will enable us to determine the lower critical dimension by mere dimensional
analysis. Our hypothesis will be validated {\it a posteriori} in the
next section, where we will derive the linearity in the context
 of an approximate maximization of the free energy.

 To first order in the
perturbative expansion in $p_1-p_2$, the variation of the free-energy
is $n \delta F=\int \d t \;  \Tr {\bf Q}\D\bQ$
computed at the unperturbed saddle point.
It is easy to see that this variation is zero for the saddle point (\ref{s0}):
\beqna
&{1\ov n}& \int \d t\; \sum_{\a\b} \bQ_{\a\b}\D\bQ_{\a\b}
\\
&=&\int \d t
 \l({\d\tp(t)\ov \d x}\r)^2
[1 - \int \d u \theta(u-u_0/2)\theta(u_0-u)-\int \d u
\theta (u-u_0/2)]=0.
\nn
\eeqna
A non zero $\delta F$ at this level would have implied, by simple dimensional
analysis, $\delta F\sim L^{D-2}$, i.e. the same result found for
ordered systems with a continuous symmetry. The  vanishing of this term
means that $D_c$ is higher than 2. One could say that fluctuations due to
the  zero mode in spin glasses cost less and become therefore important earlier
than the usual Goldstone modes in ordered systems.

 Substituting the expresion   (\ref{s0}) into the saddle points equations
 (\ref{eqt}) and denoting $u_p=3y\tp$, $\chi=3y (\d \tp/\d t)^2$ we obtain
\beqna
  {\partial^2 q(t,u)\ov\partial t^2}&=&\chi [\delta
(u-u_p)-{1\ov 2}\delta(u-u_p/2)]
\nn\\
 {\partial^2 p(t,u)\ov \partial t^2}&=&-\chi \delta
(u-u_p/2).
\label{14}
\eeqna
 We find in this way that the expresion (\ref{s0}) satisfies  the saddle point
conditions for all u
except  $u_p/2$ and
$u_p$.
It is reasonable to suppose that the effect of the Laplacian term in the
free-energy will result in a (small) smoothing of the functions $q$ and $p$
around  $u_p/2 $ and $u_p$.
Thus we suppose that the functions will have variations of a given
order $\eps$ in regions of order $\eps'$ around $u_p/2$ and $u_p$.
$\eps$ and $\eps'$, as well as $u_p$ will be in general function of $t$.
Using the monotonicity of the functional  relation between $\tp$ and $t$,
we will consider all parameters as functions of $\tp$.

We now show that under this hypothesis the following remarkable facts happen:
\begin{itemize}
\item{ $\delta F$, defined as $F(p_2,p_1)-F(p_1,p_1)$ does not depend neither
on the temperature $\tau$ nor on the magnetic field $h$.}
\item{ $\delta F$ must behave as $L^D \chi^{(5/4)}$, in order  that
the linear
interpolation (\ref{LI}) maximizes the free energy.}
\end{itemize}
Let us write the variational matrix $\bQ$ as $\bQ=\bQ^{(0)}+\delta \bQ$ where
$\bQ^{(0)}$ solves the variational problem in the absence of the perturbation.
$\delta F$ can be written as
\beqna
-2 n\delta F&=& \int \d x
\l[
   {1
  \ov 2}\ \Tr (2\de \bQ\D \bQ^{(0)}
+\de \bQ\D \de\bQ)
\r.\\
&+&\tau \Tr \de Q^2+
 \Tr \bQ^{(0)}\de \bQ^2 +
{1\ov 3}\Tr \de \bQ^3\\
&+&{y\ov 4}\l.\sum_{\a\b}(3{\bQ^{(0)}_{\a\b}}^2\de\bQ_{\a\b}^2 +4
\bQ^{(0)}_{\a\b}\de\bQ_{\a\b}^3+
\de\bQ_{\a\b}^4)
\r]
\label{dF}
\eeqna
Because of the unperturbed SPE the only terms linear in $\de \bQ$ that
can appear arise from the Laplacian term. For a given $t$ (or
equivalently $\tp$), one can evaluate the variation of the free energy
{\it density} $\de f$ just dropping the integration over $x$ in (\ref{dF}).

In this free-energy we want to study the dependence on $\tau$ and $h$
{\it before}  optimizing with respect to $\de \bQ$.
As one can see from  (\ref{s0}), the  function $p$ does not depend on $\tau$
and the function $q$ depends on $\tau$ only in the region
 $u>u_{max}$ where
 $q(u)=q_{max}$. In the same way one observes that
both functions depend on $h$ only in the region
 $u<u_{min}$ where they take the value
$q_{min}$.
To study the dependence of $\delta F$ on the temperature and the magnetic
field we
can use $q_{max}$ and $q_{min}$ as independent variables
instead of using $\tau$ and $h$
 which give rise to simpler algebra.
The structure of $\de \bQ$ we choose implies that the only terms which
 can depend on $q_{max}$ are:
\begin{itemize}
\item{$\tau \Tr \de\bQ^2$, trough its $\tau$ dependence ($\tau=q_{max}(1-3 y
q_{max}/2)$), and}
\item{$ \Tr \bQ^{(0)}\de \bQ^2$ }
\end{itemize}
The other terms do not depend on $q_{max}$ because for the $\a$ and $\b$
such that $\bQ^{(0)}_{\a\b}=q_{max}$ one has
$\de \bQ_{\a\b}=0$. So we find that
\be
-2 n {\partial \delta f\ov \partial q_{max}}=
(1-3y q_{max})\Tr \de\bQ^2+\sum_{\a\b|\bQ^{(0)}_{\a\b}=q_{max}}\de \bQ^2_{\a\b}
\label{19}
\ee
It is easy to show, using the algebra of ultrametric matrices, that
the second addendum on the r.h.s. of (\ref{19})
is exactly equal to $-(1-3y q_{max})\Tr \de\bQ^2$
and consequently $\delta f $ is independent of $\tau$.
Along the same lines one shows that the only possible dependence on $q_{min}$
is in the term $ \Tr \bQ^{(0)}\de \bQ^2$, but its derivative w.r.t. $q_{min}$
is
equal to zero.
Therefore the free-energy density variation
can only depend on $\tp$.
 Let us analize the dimensions of each term in $\delta f$
in terms of $\tp$, $\eps$, $\eps'$ and $\chi$.
We remind that while $\tp$  and $\chi$ are fixed parameter in the problem
$\eps$ and $\eps'$ are to be determined by sadlle point equations.
In the considerations which follow we
  can safely assume, that dimensionally $\eps'\sim \eps$. Keeping the two
quantities different
would only complicate the formulae, but not the scaling
of the free energy.

We find for the dimensions of the different terms in (\ref{dF}):
\beqna
 \Tr 2\de \bQ\D \bQ^{(0)} &\sim &\eps \chi
\nn\\
\Tr\de \bQ\D \de \bQ&\sim & \eps \chi \l({\partial \eps\ov\partial \tp}\r)
\nn\\
\tau \Tr \de \bQ^2  +\Tr \bQ^{(0)}\de \bQ^2   &\sim & \tp \eps^4
\nn\\
\Tr \de \bQ^3&\sim & \eps^5
\\
\sum_{\a\b}{\bQ^{(0)}_{\a\b}}^2\de\bQ_{\a\b}^2 &\sim &  \tp^2 \eps^3
\nn\\
\sum_{\a\b}\bQ^{(0)}_{\a\b}\de\bQ_{\a\b}^3&\sim & \tp \eps^4
\nn\\
\sum_{\a\b}\de\bQ_{\a\b}^4&\sim & \eps^5
\nn
\eeqna
If we now rescale:
\beqna
\eps\to \tp \eta\\
\chi\to \tp^4 \phi
\eeqna
$\delta f$ becomes an homogeneous function of order 5 in $\tp$. If the linear
form $\tp(t)$ has to be a maximum of the free-energy, the free
energy density must be independent of $\tp$. This can be seen from the fact
if one optimizes for fixed $\tp(t)$ one finds in general
$\de f=\de f(\tp,(\d\tp/dt))$. The further minimization with respect to
$\tp(t)$ leads to Euler-Lagrange equations that give zero 'acceleration'
$\d^2\tp/\d t^2$ only if $\de f$ is independent of $\tp$.
So the $\tp^5$ dependence must
be compensated at the saddle point by the dependence on $\phi\equiv
\chi/\tp^4$.
In this way, writing in all generality for the saddle point
$
\delta f= \tp^5 g(\phi)
$
one must have $g(\phi)\sim \phi^{(5/4)}\equiv \chi^{(5/4)}/\tp^5$.
Using $\chi=3y(p_2-p_1)/L$ for the form (\ref{LI}), we find that the total
free-energy variation
 scales as:
\be
\delta F\sim L^D \chi^{(5/4)}\sim |p_2-p_1|^{(5/2)}L^{D-5/2}.
\label{mr}
\ee
Equation (\ref{mr}) is the main result of this paper. It tells that
in the context of mean field theory, the lower critical dimension
at which $\delta F$ become finite for finite $|p_2-p_1|$ is $D_c=2.5$.
Moreover for $D>2.5$ one can expect fluctuations in space of the order
parameter to scale with the distance as $|q(x)-q(y)|\sim |x-y|^{1-2D/5}$.
It is worth noticing that
the value for the critical dimension we get is fully compatible
with the one found by Bray and Moore for the glassy transition
in absence of field
\cite{BMzf} with totally different methods.
The same and other authors \cite{BMf,fishuse} claimed that in any finite
dimension the spin glass transition is destroyed by
 the presence of a magnetic field.
 Our findings disagree
with this claim, as we find  $\de f$ to be  independent
on the magnetic field.

\section{A variational approximation.}
Let us now turn to an
 explicit  computation of the free-energy density increment
through a variational approach.
Instead of solving the full SPE (\ref{eqt}) we will here
 propose an explicit parametrization of the small variation to the form
(\ref{s0}) in the neighbourhood of the points $u_p/2$ and $u_p$,
 and we will {\it maximize} the free energy with
respect to the parameters of this variation. We expect
that the
numerical value so obtained for $\de F$ is
a lower bound to the real value. Furthermore we will show that it scales
as discussed in the previous section.
Thus this section proves that the leading behaviour previously obtained
does not accidentally cancel.
Moreover  here we will not need to assume the linear form (\ref{LI}) for
$\tp(t)$: this will be found as the optimum of the free energy.

We choose to smooth the singularities  around  $u_p/2$ and $u_p$
(cf. (\ref{14}))
 by  interpolating with an arc of parabola  the step-wise
linear behaviour of
the functions $q$ and $p$ in the surroundings of these points. Our variational
 functions will then be:
\be
q(u)=
\l\{
\begin{array}{ll}
q_{min} & u<u_0 \\
2u/(3y) & u_0< u<u_1 \\
\tp-a (u-u_2)^2 &  u_1<u<u_2 \\
\tp & u_2<u<u_3 \\
\tp+a' (u-u_3)^2 & u_3<u<u_4 \\
u/(3y) & u_4<u<u_5 \\
q_{max} & u>u_5
\end{array}
\r.
\ee
\be
p(u)=
\l\{
\begin{array}{ll}
q_{min} & u<u_0 \\
2u/(3y) & u_0< u<u_1 \\
\tp-a (u-u_2)^2 &  u_1<u<u_2 \\
\tp & u> u_2
\end{array}
\r.
\ee
with
\be
\begin{array}{lll}
u_0=3y q_{min}/2 &
u_1=3 y \tp/2-\delta/2 &
u_2=3 y \tp/2+\delta/2 \\
u_3=3 y \tp-\delta'/2 &
u_4=3 y \tp+\delta'/2 &
u_5=3 y q_{max} \\
a=1/(3y \delta ) &
a'=1/(6 y \delta'). &
\end{array}
\ee
The reader should not be confused by the notation at this point,
although $q_{min}$ and $q_{max}$ are the same as in the presious sections
and of (\ref{s0}), we changed here the notation for the points $u_i$
($i=0,...,5$). The only  variational parameters that  appear in the
the free energy functional are
$\delta$ and $\delta'$. Denoting as in the previous
section  $\chi=3 y\l({\d \tp\ov \d t}\r)^2$,
the  free energy density increment as a function of $\de$ and $\de'$ takes the
form:
\be
\delta f={{31 {{\delta }^5}}\over {102060 {y^3}}} -
   {{{{\delta }^4} {\it \tp}}\over {324 {y^2}}} +
   {{\chi  \delta }\over {9 y}} + {{\chi  \delta '}\over {18 y}} -
   {{{{\delta }^3} {{\delta '}^2}}\over {9720 {y^3}}} +
   {{{{\delta }^2} {\it \tp} {{\delta '}^2}}\over {648 {y^2}}} -
   {{{\it \tp} {{\delta '}^4}}\over {5184 {y^2}}} -
   {{11 {{\delta '}^5}}\over {3265920 {y^3}}}
\label{feee}
\ee
which has to be maximized with respect to $\de$ and $\de'$.
Equation (\ref{feee}) is consistent with the scaling established
in the previous section.
The change of variables:
\beqna
\de &=&y \tp {a+b\ov 2}
\nn\\
\de'& =&y \tp { b-a\ov 4}\\
\chi &=&y^2 \tp^4 \phi
\nn
\eeqna
gives us
\beqna
\delta f&=&{y^2 \tp^5}{b\ov 52254720}
\left( 29\,{a^4} - 10080\,{a^2}\,b - 252\,{a^3}\,b
\r.
\nn
\\
& + &
 \l. 142\,{a^2}\,{b^2} - 84\,a\,{b^3} - 11\,{b^4} + 2903040\,\phi  \right).
\eeqna
It is apparent form the
variational equations for $a$ and $b$ that
 $a$ should be of order $\phi^{1/2}$ while
$b\simeq \phi^{1/4}$. To lowest order in $\phi$ the solution
is $a=- b^2 /240$, $b=12 \sqrt{2} (7/11)^{(1/4)}\phi^{1/4}$
which gives for the free-energy density
\be
\delta f=0.673659 \ y^2 \chi^{5/4}.
\ee
This result confirms in a specific example the behaviour in $\chi^{5/4}$
of $\de f$ which is the only one compatible with $\tp(t)$ linear in $t$.
The total free-energy of the interface,
$\delta F=\delta f L^D $ is proportional to $L^{D-5/2} |p_1-p_2|^{5/2}$,
confirming the analysis of the previous section.
Let us observe here the proportionality of $\de F$ to $y^2$,
the coefficient of the quatic term in (\ref{landau}).

\section{Conclusions}
In this paper we have shown that it is possible to estimate
the cost in free energy of a domain wall between two different phases in
spin glasses.
This information was used to indicate a possible value for the
lower critical dimension $D_c=2.5$ for replica symmetry breaking.

We are aware of the several
criticisms that could be raised against this indication.
In ordered system MFT gives a reliable estimate of $D_c$ because the
fluctuation of the order parameter are negligible at very low temperature.
In the spin glass case there is not such evidence.
Moreover, in ordinary systems, this kind of analysis
is confirmed
by the behaviour of the perturbation expansion. In $O(N)$ models,
for example, the free propagator $G(k)\sim k^{-2}$
in the ordered phase and therefore the fluctuations
of the order parameter diverge in
$D=2$. In the spin glass case, there is a whole family of propagators, the
most divergent of which, in presence of magnetic field, has the behaviour
$G(k)\sim k^{-3}$ \cite{dedoko}.
If this behaviour is not modified by
renormalization it would imply a lower critical dimension $D_c=3$.

At present we do not know if any of this criticisms will be substantiated
and the actual critical dimension in spin glasses is
larger then 2.5. But if this happens, the replica symmetry restoration
mechanism must be different from the simple
'instantonic' one that we have proposed in this paper.

The present state of the art in numerical simulations of spin glasses
indicates that in dimension two there is no transition \cite{sim2D} while in
dimension
four there is full replica symmetry breaking \cite{sim4D}.
Unfortunatly the situation is
far from clear in dimension three. Recent simulations  by one of us
\cite{sim3D} obtained from large
lattices were compatible with a finite temperature phase
 transition
but also with
an essential singularities at
$T=0$, which would  indicate $D_c=3$.

All this calls for further research.
One can expect that simulations on larger lattices will eventually resolve
the problem of the existence of the transition
in three dimensions.
On the more analytical side a crucial problem to attack is the
renormalization of the $k^{-3}$ behaviour of the most singular propagator
in perturbation theory.

\section*{Acknowledgments}
It is a pleasure to
thank P.W.Anderson, C. de Dominicis, M.M\'ezard, D.Sherrington
for interesting discussions.

\end{document}